\documentclass[pre,onecolumn]{revtex4}
\usepackage{graphicx}
\begin{document}
\title{Spatially-Periodic Cluster Pattern of Coupled Forced Oscillators}
\author{Hidetsugu Sakaguchi}
\address{Interdisciplinary Graduate School of Engineering Sciences, \\
Kyushu University, Kasuga, Fukuoka 816-8580, Japan}
\begin{abstract}
We propose a simple model for periodic clustering of particles under forced oscillation. Effective viscosity is assumed to increase owing to neighboring particles by analogy with the Einstein viscosity law. The linear stability analysis and numerical simulations show that the uniform distribution is unstable, and spatially-periodic and stripe patterns appear respectively in one and two dimensions.\end{abstract}
\maketitle

Pattern formation has been intensively studied as dissipative structure in nonequilibrium systems~\cite{Cross}. There are various mechanisms of pattern formation. The Turing instability is a mechanism of pattern formation in reaction-diffusion equations.  Faraday waves are standing wave patterns on liquids and granular materials formed owing to the parametric instability by perpendicular oscillation~\cite{Melo}. In some cases, patterns appear through the clustering of particles. Wind ripple~\cite{Bagnold,Werner} and sand ripple~\cite{Blondeaux} are stripe patterns formed in granular material, where the interaction of granular particles and the surrounding fluid is important. In Kundt's experiment to visualize a sound wave, a periodic pattern of cork powder called striations or vertical curtains is observed over the standing wave pattern with the wavelength of the sound wave. The wavelength of the striations is much smaller than the wavelength of the sound wave. The striations have been studied for a long time, beginning with the study of acoustic streaming by Andrade~\cite{Andrade}. The uniform state of particle density becomes spontaneously unstable under the external forcing; however, the formation mechanism of the short-wavelength pattern is not completely understood.

We consider a mechanism of pattern formation owing to the clustering of particles induced by external forcing. 
Einstein's law of viscosity expresses that the viscosity of fluids, including spherical small particles, increases as $1+2.5\phi$ where $\phi$ is the volume fraction of the particles. By analogy with Einstein's viscosity law, we consider a simple model of coupled forced oscillators where the effective damping coefficient increases due to the interaction with the surrounding particles. 
A one-dimensional model equation is written as    
\begin{equation}
\frac{d^2x_i}{dt^2}=F\sin\omega t-\alpha\sum_{i^{\prime}}e^{-(x_{i^{\prime}}-x_i)^2/D}\frac{dx_i}{dt},
\end{equation}
where $F\sin\omega t$ denotes the external forcing with frequency $\omega$ and $\alpha$ is the damping coefficient. The term $\sum_{i^{\prime}}\exp\{-(x_{i^{\prime}}-x_i)^2/D\}$ represents the effective increase of the damping coefficient due to the interaction with neighboring particles, where the summation is taken for all particles including $i$, and $D$ determines the range of interaction. 
When the particles are placed at a constant interval $d$, the effective viscosity is expressed as
\[\beta=\alpha(1+2\sum_{m=1}^{\infty} e^{-m^2d^2/D}).\] 
The effect of initial conditions decays with time in forced oscillation with damping. All particles oscillate in sync sufficiently after the decay time as
$x_i(t)=d(i-1/2)+x_v(t)$ where $x_v(t)=A\sin\omega t+B\cos\omega t$ with 
$A=-F/(\omega^2+\beta^2)$ and $B=-F\beta/\{\omega(\omega^2+\beta^2)\}$. 

To investigate the stability of the uniform arrangement, we perform the linear stability analysis. 
The perturbations $\delta x_i(t)$ obey
\begin{equation}
\frac{d^2\delta x_i}{dt^2}=-\beta\frac{d\delta x_i}{dt}-\alpha\frac{dx_v}{dt}\sum_{i^{\prime}\ne i}e^{-(i^{\prime}-i)^2d^2/D}\{-2(i^{\prime}-i)d/D\}(\delta x_{i^{\prime}}-\delta x_i).
\end{equation}
If $\delta x_i$ is expressed as $\delta x_i=\sum_k\{x_{ck}\cos(kdi)+x_{sk}\sin(kdi)\}$ by the Fourier transform, $x_{ck}$ and $x_{sk}$ obey
\begin{equation}
\frac{d^2x_{ck}}{dt^2}=-\beta \frac{dx_{ck}}{dt}-\gamma_k\frac{dx_v}{dt}x_{sk},\;\;
\frac{d^2x_{sk}}{dt^2}=-\beta \frac{dx_{sk}}{dt}+\gamma_k\frac{dx_v}{dt}x_{ck},
\end{equation}
where
\[\gamma_k=\alpha\sum_{m=1}^{\infty}e^{-m^2d^2/D}(4md/D)\sin (kdm).\]
The Mathieu equation for the parametric resonance of a swing is the second-order ordinary differential equation with a time-periodic coefficient. Equation (3) has the form of coupled Mathieu equations with damping terms. We solve the solution numerically for one period $T=2\pi/\omega$ staring from an initial conditions $x_{ck}=1$, $x_{sk}=dx_{ck}/dt=dx_{sk}/dt=0$, and $x_{ck}(T)$, $x_{sk}(T)$, $dx_{ck}(T)/dt$, and $dx_{sk}(T)/dt$ are set to be $r_{11}$, $r_{21}$, $r_{31}$, and $r_{41}$, respectively. Similarly, we obtain $r_{12}$, $r_{22}$, $r_{32}$, $r_{42}$ starting from $x_{sk}=1$, $x_{ck}=dx_{ck}/dt=dx_{sk}/dt=0$, $r_{13}$, $r_{23}$, $r_{33}$, $r_{43}$ starting from $dx_{ck}/dt=1$, $x_{ck}=x_{sk}=dx_{sk}/dt=0$, and $r_{14}$, $r_{24}$, $r_{34}$, $r_{44}$ staring from d$x_{sk}/dt=1$, $x_{ck}=x_{sk}=dx_{ck}/dt=0$. The eigenvalues of the $4\times 4$ matrix $R$ whose $(i,j)$ component is $r_{i,j}$. determines the stability of Eq.~(3).

Figure 1(a) shows $\lambda=(\ln \lambda_R)/T$ where $\lambda_R$ is the largest eigenvalue of the matrix $R$ as a function of wavenumber $k$ for $D=0.1$ and 0.25 at $d=0.05$, $\alpha=0.05$, $F=1$, and $\omega=3$.  
$\lambda(k)$ is always positive and has a peak at $k=4.45$ for $D=0.1$ and $k=2.95$ for $D=0.25$.  Direct numerical simulation was performed in a system of size $L=10$ with periodic boundary conditions. The particle number is $N=200$, and the initial regular interval is $d=L/N=0.05$. The initial conditions are $x_i(0)=(i-1/2)d+\delta x_i(0)$, where $\delta x_i(0)$ is set by a uniform random number between $[-0.000025,0.000025]$.    
Figure 1(b) shows the time evolution of $x_i$ for $i=1,2,\cdots, N$ at $D=0.25$. Clustering occurs with time, and six clusters appear at $t=600$. Figure 1(c) shows the time evolution of $x_i$'s for $D=0.1$. Eight clusters appear at $t=600$.  The cluster number did not change after $t=600$. Figure 1(d) shows the relationship between $i$ and $x_i$ at $t=5000$ for $D=0.1$. In each cluster, $x_i$ takes almost the same position. The average wavenumbers for the $n$ clusters is $2\pi n/L$, which are 3.77 for $D=0.25$ and  5.02 for $D=0.1$. The average wavenumbers are close to the wavenumber for the maximum of $\lambda(k)$ in Fig.~1(a). That is, the spatially-periodic clustering is approximately understood by the linear stability analysis.

\begin{figure}[h]
\begin{center}
\includegraphics[height=3.5cm]{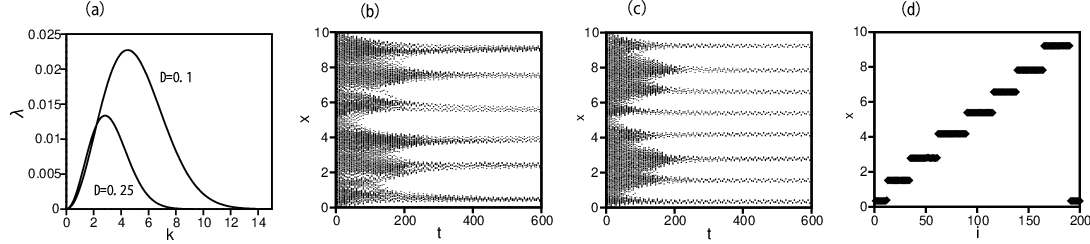}
\end{center}
\caption{(a) Relationship between $k$ and $\lambda(k)$ for $D=0.25$ and 0.1. (b) Time evolution of $x_i$'s for $D=0.25$. (c) Time evolution of $x_i$'s for $D=0.1$. (d) Snapshot of $x_i$ for $i=1,2, \cdots,200$ at $t=5000$ for $D=0.1$.}
\label{f1}
\end{figure}

The one-dimensional model can be generalized to two dimensions. Initially, particles are uniformly arranged in a rectangular space of $L_x\times L_y$, i.e., the particles are placed at the interval $d$ both in the $x$ and $y$ directions. The positions in the $y$ direction are assumed to be constant in time, that is, each particle on the line $y=(j-1/2)d$ moves only in the $x$ direction. The equation of motion for $x_{i,j}$ is expressed as
\begin{equation}
\frac{d^2x_{i,j}}{dt^2}=F\sin\omega t-\alpha \sum_{i^{\prime}}\sum_{j^{\prime}}e^{-\{(x_{i^{\prime},j^{\prime}}-x_{i,j})^2+(j^{\prime}-j)^2d^2\}/D}\frac{dx_{i,j}}{dt}.
\end{equation}
\begin{figure}[h]
\begin{center}
\includegraphics[height=4.5cm]{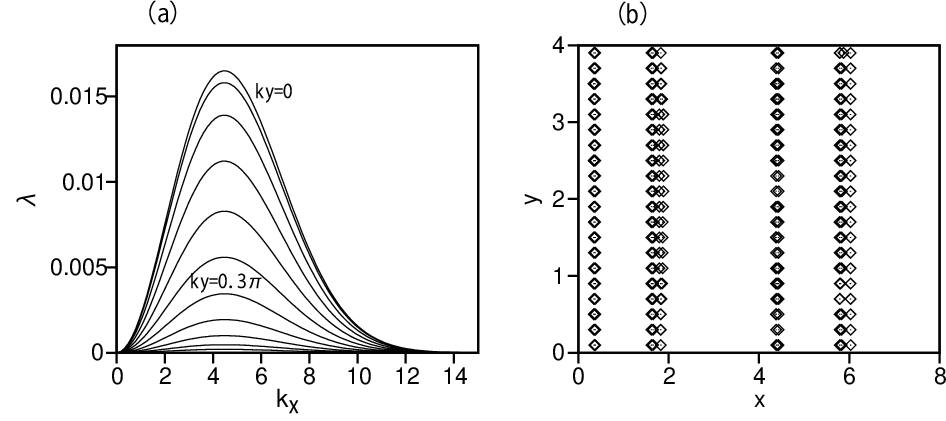}
\end{center}
\caption{(a) Relationship between $k$ and $\lambda(k_x)$ for $k_y=6\pi/100\times l$ for $l=0,1,2, \cdots,10$. System parameters are $F=1$, $\omega=3$, $d=0.2$, $D=0.1$, and $\alpha=0.05$. (b) Snapshot of $x_{i,j}$ for $i=1,2,\cdots,40$ and $j=1,2, \cdots,20$ at $t=5000$.}
\label{f1}
\end{figure}
There is a uniformly oscillating solution of the form $x_{i,j}=(i-1/2)d+x_v(t)$ where $x_v=A\sin\omega t+B\cos\omega t$ with $A=-F/(\omega^2+\beta^2)$ and $B=-F\beta/\{\omega(\omega^2+\beta^2)\}$. In the two-dimensional model, $\beta$ is expressed as 
\[\beta=\alpha\left (1+4\sum_{m_x=1}^{\infty} e^{-m_x^2d^2/D}+4\sum_{m_y=1}^{\infty}\sum_{m_x=1}^{\infty} e^{-(m_x^2d^2+m_y^2d^2)/D}\right ).\]   
If the perturbation $\delta x_{i,j}(t)$ is expanded 
as $\delta x_{i,j}=\sum_{k_y}\sum_{k_x}\{x_{ck_xk_y}\cos(k_xdi)+x_{sk_xk_y}\sin(k_xdi)\}\cos(k_ydj)$ by the Fourier transform, $x_{ck_xk_y}$ and $x_{sk_xk_y}$ obey
\begin{equation}
\frac{d^2x_{ck_xk_y}}{dt^2}=-\beta \frac{dx_{ck_xk_y}}{dt}-\gamma_{k_xk_y}\frac{dx_v}{dt}x_{sk_xk_y},\;\;
\frac{d^2x_{sk_xk_y}}{dt^2}=-\beta \frac{dx_{sk_xk_y}}{dt}+\gamma_{k_xk_y}\frac{dx_v}{dt}x_{ck_xk_y},
\end{equation}
where
\[\gamma_{k_xk_y}=\alpha\left (\sum_{m_x=1}^{\infty}e^{-m_x^2d^2/D}(4m_xd/D)\sin (k_xdm_x)\right )\left (1+2\sum_{m_y=1}^{\infty}e^{-m_y^2d^2/D}\cos(k_ydm_y)\right ).\]

The linear growth rate of the perturbation can be calculated similarly to the one-dimensional model for $(k_x,k_y)$. Figure 2(a) shows $\lambda(k_x,k_y)$ at $k_y=6\pi/100\times l$ for $l=0,1,2,\cdots,10$. The curve of $\lambda(k_x)$ at the top corresponds to $k_y=0$, and the lower curves correspond to larger $k_y$.    
The parameters are $F=1$, $\omega=3$, $d=0.2$, $D=0.1$, and $\alpha=0.05$.  The growth rate $\lambda(k_x)$ takes a peak value at $k_x=4.4$ for $k_y=0$. The peak value of the growth rate decreases with $k_y$.  Direct numerical simulation was performed in a rectangular system of $L_x\times L_y=8\times 4$. There are $40$ particles on the line $y=d(j-1/2)$ for $j=1,2,\cdots,20$. The initial condition is $x_{i,j}=d\{(i-1/2)+0.0001r_{i,j}\}$ where $r_{i,j}$ is a uniform random number between $[-0.5,0.5]$. Figure 2(b) shows a snapshot of $x_{i,j}$ at $t=5000$. Roughly, four linear clusters are formed, although there is fine splitting in the second and fourth clusters. The particles align linearly in the $y$ direction, because $\lambda(k_x,k_y)$ takes the maximum at $k_y=0$. 

To summarize, we have proposed a simple phenomenological model in which the effective viscosity increases due to the interaction with neighboring particles. Spatially-periodic patterns appear due to the instability of the uniformly oscillating state. The spatially-periodic cluster pattern due to the forced oscillation might be related to the mechanism of pattern formation of the striations in Kundt's tube experiment. Another possibly related phenomenon is the memory effect of drying pastes, that is, cracks appear in the perpendicular direction of the vibration before desiccation. Recently, Kitsunezaki et al. observed by the X-ray tomography that powder particles take an anisotropic arrangement relative to the direction of the vibration~\cite{Nakahara}.


\begin{thebibliography}{99}
\bibitem{Cross} M.~C.~Cross and P.~C.~Hohenberg, Rev. Mod. Phys. {\bf 65}, 851 (1993).
\bibitem{Melo} F.~Melo, P.~Umbanhowar, and H.~L.~Swinney, Phys. Rev. Letts. {\bf 72}, 172 (1994).
\bibitem{Bagnold} R.~A.~Bagnold {\it The Physics of Blown Sand Desert Dune} (Chapman and Hall, London, 1954).
\bibitem{Werner} B.~T.~Werner and D.~T.~Gillespie, Phys. Rev. Letts. {\bf 71}, 3230 (1993).
\bibitem{Blondeaux} P.~Blondeaux, J. Fluid. Mech. {\bf 218}, 1 (1990).
\bibitem{Andrade} E.~N.~da C.~Andrade, Proc. R. Soc. A, {\bf 134}, 445 (1931).
\bibitem{Nakahara} S.~Kitsunezaki, A.~Nishimoto, T.~Mizuguchi, Y.~Matsuo, A.~Nakahara, Phys. Rev. E {\bf 105}, 044902 (2022).
\end{thebibliography}
\end{document}